\documentclass[12pt,preprint]{aastex}
\begin{document}

\parindent=1.0cm

\title{The Distribution of Main Sequence and Pre-Main Sequence Stars in the Young 
Anticenter Cluster NGC 2401 \altaffilmark{1}\altaffilmark{2}\altaffilmark{3}}

\author{T. J. Davidge}

\affil{Dominion Astrophysical Observatory,
\\National Research Council of Canada, 5071 West Saanich Road,
\\Victoria, BC Canada V9E 2E7\\tim.davidge@nrc.ca}

\altaffiltext{1}{Based on observations obtained at the Gemini Observatory, which is
operated by the Association of Universities for Research in Astronomy, Inc., under a
cooperative agreement with the NSF on behalf of the Gemini partnership: the National
Science Foundation (United States), the National Research Council (Canada), CONICYT
(Chile), the Australian Research Council (Australia), Minist\'{e}rio da Ci\^{e}ncia,
Tecnologia e Inova\c{c}\~{a}o (Brazil) and Ministerio de Ciencia, Tecnolog\'{i}a e
Innovaci\'{o}n Productiva (Argentina).}

\altaffiltext{2}{This research used the facilities of the Canadian Astronomy Data
Centre operated by the National Research Council of Canada with the support
of the Canadian Space Agency.}

\altaffiltext{3}{This research has made use of the NASA/IPAC Infrared Science Archive,
which is operated by the Jet Propulsion Laboratory, California Institute of Technology,
under contract with the National Aeronautics and Space Administration.}

\begin{abstract}

	Images obtained with the Gemini Multi-Object Spectrograph on Gemini South are 
used to examine the photometric properties and spatial distributions of main sequence 
(MS) and pre-main sequence (PMS) objects in the star cluster NGC 2401. The data sample 
several magnitudes fainter than previous studies, and a large population of candidate 
PMS (cPMS) stars are identified. The cPMS stars are traced out to 2.4 arcmin from the 
cluster center, and have a flatter spatial distribution 
than the brightest MS stars near the cluster center. The luminosity function of 
all MS and candidate PMS stars can be matched by a model that assumes a solar 
neighborhood mass function, suggesting that NGC 2401 has not yet shed significant 
numbers of members with masses $\geq 0.5$ M$_{\odot}$. The frequency of wide binaries 
among the MS stars is $\sim 3\times$ higher than among the cPMS stars. It is argued that 
the difference in the spatial distributions of MS and PMS objects is not the consequence 
of secular dynamical evolution or structural evolution driven by 
near-catastrophic mass loss. Rather, it is suggested that the 
different spatial distributions of these objects is the fossil imprint of primordial 
sub-clustering that arises naturally if massive stars form preferentially 
in the highest density central regions of a protocluster.

\end{abstract}


\section{INTRODUCTION}

	While there is a large body of evidence suggesting that stars form in 
groups rather than in isolation, the vast majority of stars in the 
Galaxy and neighboring systems are not in obvious 
clusters or associations (e.g. Lada \& Lada 2003; Bonatto \& Bica 2011; 
Silva-Villa \& Larsen 2011). The seeming inconsistency between the unclustered 
nature of stars and the presence of highly organized natal environments 
can be reconciled if the majority of clusters are disrupted early in 
their lives. Age $vs.$ number relations of clusters in external galaxies investigated by 
Fall \& Chandar (2012) suggest a disruption rate $\sim 1$ dex per decade in age. 
The diverse nature of the galaxies considered by Fall \& Chandar (2012) 
suggest that the pace of cluster disruption is not sensitive to global conditions. 
This being said, detailed investigations of cluster statistics within galaxies 
suggest that the rate of cluster disruption may depend on environment 
(e.g. Bastian et al. 2011; de Grijs et al. 2013; Silva-Villa et al. 2014). 

	Even if the majority of clusters survive for only a few Myr, 
this is still sufficient time for the cluster environment to influence the properties of 
member stars. For example, the radiation field from massive stars may erode 
accretion disks around pre-main sequence (PMS) objects (e.g. Johnstone et 
al. 1998; Adams et al. 2004; De Marchi et al. 2011), thereby choking subsequent growth 
and affecting the mass function (MF) of stars that ultimately leave the cluster. 
The MF in environments that contain a large population of hot, young stars might then 
contain a higher fraction of low mass stars than in the Solar 
Neighborhood. Dynamical interactions within the cluster can 
also affect the binary frequency through the preferential disruption of binaries 
that contain low mass companions (e.g. Marks \& Kroupa 2012).

	Young star clusters are important laboratories for probing 
cluster evolution. NGC 2401 is a young open cluster in the Perseus 
spiral arm. It has been the subject of four recent photometric studies, and the 
ages, distances, and reddenings obtained in these are summarized in Table 1. 
Sujatha et al. (2004), Baume et al. (2006), and Hasegawa, et al. (2008) estimate the 
age of NGC 2401 from the main sequence turn-off (MSTO), while 
Davidge (2014) measure an age using the faintest MS stars, 
which define the MS cut-off (MSCO). The dispersion in the ages found from these 
studies are almost certainly linked to the problematic nature of distinguishing between 
cluster and non-cluster members in the NGC 2401 CMD.

	Davidge (2014) investigated the MF and 
spatial distribution of stars in NGC 2401. The luminosity function (LF) of 
the center of NGC 2401 in $K$ was matched by a model that assumes a 
Chabrier (2001) MF. Baume et al. (2006) also concluded that stars 
more massive than solar in NGC 2401 follow a solar neighborhood-like MF. 
In the present study, deep $g'$ and $i'$ images obtained with the 
Gemini Multi-Object Spectrograph (GMOS) on Gemini South are used to examine 
the photometric properties and spatial distributions of MS 
and PMS objects in NGC 2401. These data probe much deeper than previous 
studies and have sub-arcsec image quality, thereby allowing sources to be resolved 
that might be mis-identified as a single object during poorer seeing conditions. 
A population of faint candidate PMS (cPMS) stars 
are identified. While it is not possible to identify with confidence individual PMS stars 
with the existing photometry alone due to field star contamination, 
it is still possible to investigate in a a statistical manner the basic properties of the 
cPMS stars as a group, such as their number density and distribution on the sky. 

	The paper is structured as follows. Details of the observations, the 
steps used to reduce the images, and the procedures used to make the photometric 
measurements are presented in Section 2. The color-magnitude diagram (CMD) of NGC 2401 
and the identification of cPMS stars are the subject of Section 3. In Section 4 it is 
shown that the distribution of cPMS stars on the sky is indicative of 
cluster membership for a large fraction of these, although it is also shown that MS 
stars and the cPMS objects have systematically different spatial distributions, 
in the sense that the cPMS stars have a flatter radial distribution near the cluster 
center than the MS stars. In Section 5, and it is shown that the cluster LF follows a 
solar neighborhood-like relation after correcting statistically for field stars. 
A summary and discussion of the results follows in Section 6.

\section{OBSERVATIONS \& REDUCTIONS}

\subsection{Description of the Observations}

	NGC 2401 was observed in $g'$ and $i'$ with GMOS (Hook et al. 2004) on Gemini 
South during the night of December 29, 2013 as part of program GS2014A-Q-84 (PI: 
Davidge). The detector at that time was a mosaic of three EEV $2048 \times 6048$ CCDs, 
with each $13.5\mu$m pixel sampling 0.073 arcsec on a side \footnote[4]{The EEV CCDs have 
since been replaced with Hamamatsu devices.}. The images were binned 
$2 \times 2$ during read-out. The light profiles of isolated stars have 
a full-width at half maximum of 0.5 arcsec FWHM in $i'$ and 0.6 arcsec in $g'$.
The sky conditions were photometric when the data were recorded. 

	Photometry of stars with $i' > 17$ was obtained from a single 
200.5 sec exposure in $g'$, and five 40.5 sec exposures in $i'$. 
A single 1.5 sec exposure was also recorded in each filter so that 
photometric measurements could be made of stars with $i' < 17$. 
The bright limit of the short exposure images is $g' \sim i' 
\sim 15$, which is below the MSTO. 

\subsection{Data Reduction and Photometric Measurements}

	The first step in the processing of the images was bias subtraction. A series of 
bias frames with $2 \times 2$ binning were recorded on December 29, 2013. 
A master bias frame was constructed from these by taking the median intensity on a 
pixel-by-pixel basis, and the result was subtracted from the raw images.

	Flat field frames with $2 \times 2$ binning were obtained from a series 
of $g'$ and $i'$ twilight sky exposures that were recorded on December 28, 2013. 
Final flat-field frames were constructed by taking the median signal 
at each pixel location after the individual twilight sky observations had been 
normalized to unity, and the bias-subtracted images were divided by these. Finally, 
an $i'$ fringe frame that was constructed from data that were taken in November 2013 was 
subtracted from the flat-fielded $i'$ images. The deep 
fringe-corrected $i'$ exposures were aligned, and the 
results were combined by taking the median flux at each pixel after 
correcting for exposure-to-exposure differences in the mean sky level.

	Stellar brightnesses were measured with the point spread function 
(PSF)--fitting program ALLSTAR (Stetson \& Harris 1988). The source 
catalogues, initial brightness estimates, and PSFs used by ALLSTAR were obtained 
by running the appropriate tasks in DAOPHOT (Stetson 1987). At least 30 unsaturated and 
isolated stars were combined to construct a PSF for each filter $+$ exposure time 
pair. Faint companions were removed by subtracting them from the images
using progressively improved versions of the PSFs.

	The photometric calibration was set using zeropoints 
obtained from observations of Smith et al. (2002) standards that were 
recorded in January 2014. That the standards were not recorded on the same night 
as the science data introduces uncertainties in the calibration. 
Based on the compilation of GMOS photometric zeropoints discussed by Jorgensen (2009), 
and given that the primary mirror was not washed or re-coated between the time when 
the data were recorded and the nights that the standards were observed, then 
the uncertainty in the calibration of each filter is 
roughly $\pm 0.05$ magnitude. However, correlations in 
filter-to-filter departures from mean trends are such that the uncertainty in color 
measurements is smaller than those in the individual magnitude measurements 
(e.g. Figure 3 of Jorgensen 2009).

	Artificial star experiments were run to assess completeness and estimate the 
random errors in the photometry. Preliminary experiments indicated that the random errors 
in the measurements with $i' < 20$ were small, and so artificial stars were assigned 
magnitudes and colors that matched those of the faint red sequence in the CMDs, 
which has $g'-i' \sim 3$ and is identified as containing cPMS stars in Section 3. 
Only stars that were successfully recovered in both filters were considered to be 
detected for the purposes of calculating completeness. These experiments indicate that 
the GMOS photometry of cPMS objects is at least 90\% complete when $i' < 21.5 - 22$.

\section{A DEEP COLOR-MAGNITUDE DIAGRAM OF NGC 2401}

	The $(i', g'-i')$ CMDs of objects in different parts of the GMOS field 
are shown in Figure 1. Regions 1 -- 5 sample equal areas on the sky and 
are centered on the main concentration of bright stars in NGC 2401. The 
photometric measurements of objects with $i' < 17$ are from the 1.5 sec 
exposures, and detector saturation defines the bright limit. 
There are 542, 362, and 334 objects in the CMDs of Regions 1, 2, and 3. 
The CMDs of Regions 4 and 5 contain statistically similar numbers 
of objects (276 in Region 4 and 284 in Region 5), suggesting that they 
sample an area where there are few -- if any -- cluster stars. 
The boundaries of the five regions on the GMOS $g'$ image are shown in Figure 2.

	The CMDs in Figure 1 sample objects that are many magnitudes fainter 
than those studied by Baume et al. (2006), and a rich population of faint cPMS stars 
is detected (see below). The GMOS CMD of objects in Regions 1, 2, and 3 is compared 
with the CMD of sources in the same area but based 
on the Baume et al. (2006) photometry in Figure 3. 
The Baume et al. (2006) measurements were transformed into the SDSS photometric system 
using the relations in Table 7 of Smith et al. (2002), and so only sources with 
measurements in B, V, R, and I are included. The GMOS and Baume et al. (2006) CMDs 
in Figure 3 are very similar in the overlapping magnitude range. The $g'-i'$ colors 
obtained from the GMOS observations and the $V-I$ colors measured by Baume et al. (2006) 
are compared in the Appendix. 

	NGC 2401 is at a low Galactic latitude, and there is substantial 
contamination from non-cluster sources. Baume et al. (2006) sample stars over a 
larger field than was covered by GMOS, and their data thus may provide insights into 
field star contamination in the upper portions of the GMOS CMDs. The $(i', g'-i')$ CMD 
of stars in the Baume et al. (2006) dataset that are external to 
Region 3, which was transformed into the SDSS photometric system using the procedure 
described above, is shown in the right hand panel of Figure 3. 
There is a prominent near-vertical sequence with $g'-i'$ between 
0.5 and 1.5 that spans a wide range of brightnesses. This 
sequence is steeper than what would be expected from MS stars 
at the distance of NGC 2401, and is likely populated by low mass foreground stars.
This sequence is seen in the composite GMOS Region 1, 2, and 3 CMD in the left hand panel 
of Figure 3, although it is not well-defined given the small area that is 
covered and the presence of stars that belong to NGC 2401. 

	The composite CMDs of Regions 1, 2, and 3 (left hand panel) and 
Regions 4 and 5 (right hand panel) are compared with isochrones 
from Bressan et al. (2012) in Figure 4. As the GMOS data do not 
sample the upper portions of the cluster MS, photometric measurements of sources 
from Baume et al. (2006) that have $V < 17$ and are in their `cluster area' are also 
plotted in Figure 4. The Baume et al. (2006) measurements were transformed 
into the SDSS photometric system using relations from Smith et al. (2002). 

	The error bars in the left hand panel of Figure 4 show 
the $\pm 2\sigma$ dispersions in the $g'-i'$ colors of stars with $i' > 22$ and $g'-i' 
\sim 3$ that were obtained from the artificial star experiments. At faint 
magnitudes the uncertainties obtained from the artificial star experiments are 
larger than those predicted from the fitting errors computed by 
ALLSTAR. The artificial star experiments predict dispersions in $g'-i'$ of $\pm 0.05$ 
magnitude at $i' = 21$ and $\pm 0.14$ magnitude at $i' = 22$. For comparison, 
the PSF fitting errors computed by ALLSTAR predict dispersions of $\pm 0.03$ 
magnitudes at $i' = 21$ and $\pm 0.07$ magnitudes at $i' = 22$.

	The MS of NGC 2401 is clearly seen in the composite Region 1 -- 3 CMD in Figure 
4 when $i' < 17.5$. In fact, the difference in the number of 
sources with $g'-i'$ between 0.5 and 1.0 and $i'$ between 17 and 17.5 
in Regions 1, 2, and 3 and Regions 4 $+$ 5 is $22 \pm 7$. However, in 
the same color range but 0.5 magnitudes in $i'$ fainter 
(i.e. $i'$ between 17.5 and 18) the difference drops to 
only $11 \pm 9$ objects. That a well-defined population of MS is not found at 
magnitudes $i' \geq 17.5$ suggests that members of NGC 2401 that are fainter than 
$i' \sim 17.5$ have not yet evolved onto the MS, and this is consistent with an 
age of a few tens of Myr for NGC 2401.

	The isochrones in Figure 4 were constructed from models with Z=0.020, which 
is the metallicity measured for the Pleiades by Soderblom 
et al. (2009). $E(B-V) = 0.36$ from Baume et al. (2006) has been adopted, while 
A$_{i'}$ and $E(g'-i')$ were calculated assuming 
A$_{i'} = 2.086 \times E(B-V)$ and $E(g'-i') = 1.707 \times E(B-V)$. 
These relations are from Table 6 of Schegel et al. (1998), which in turn 
are based on the R$_V = 3.1$ model of Cardelli et al. (1989). The adopted reddening 
produces good agreement between the isochrones and the blue envelope of the MS. 

	A distance modulus of 13.6 was adopted for the comparisons in Figure 4. The mean 
distance modulus computed from the entries in Table 1, not including that 
found by Sujatha et al. (2004), which is substantially smaller than the other 
three, is 13.8. The distance modulus of 14.0 found by Baume et al. (2006) 
does not produce good agreement between the isochrones and the observations. 

	It can be seen from Figure 4 that if NGC 2401 had an age 
older than $\sim 50$ Myr then there should be a well-defined 
MS extending to $i' \geq 20$. The lack of a MS at these 
magnitudes is not due to sample incompleteness, as the artificial 
star experiments indicate that completeness only becomes an issue at much fainter 
magnitudes (Section 2.2). However, there is a population of sources with $i' > 21$ 
that are concentrated near $g'-i' \sim 3$ in the composite Region 1 -- 3 CMD. Objects 
with similar photometric properties are also seen in Regions 4 and 5, although there is 
a clear excess number in the Region 1 -- 3 CMD. 
In Sections 4 and 5 it is demonstrated that objects with $i' > 21$ 
in Regions 1 -- 3 are present in numbers that are significant at the many sigma level 
after correcting statistically for contamination from field objects, and that the 
projected density of these objects on the sky increases with decreasing distance from 
the cluster center. There is also a clear tendency for the concentration of objects at 
faint magnitudes in the CMD to fall redward of the MS expected for an older population. 
Given the lack of faint cluster MS stars, coupled with the spatial distribution and 
red color of the faint objects on the CMD, then they are identified as cPMS stars. 

	The locus of faint red objects in the left 
hand panel of Figure 4 is not reproduced by the isochrones. The 10 Myr isochrone comes 
closest to matching the colors of the objects having $g'-i' > 3$ near $i' \sim 22$, 
although this model only skirts the blue envelope of the faint red concentration.
Previous studies have had difficulties reproducing the photometric 
properties of PMS stars at visible wavelengths (e.g. Lyra et al. 2006; Bell et al. 2012).
PMS stars with ages near 30 Myr may retain accretion disks (e.g. De Marchi et al. 2011; 
2013) that will be a source of circumstellar extinction, and hence redden visible colors. 
The inability to match photometric properties at visible wavelengths may also be tied to 
the presence of spots on PMS objects that are related to accretion activity (e.g. 
Stauffer et al. 2003; Pecaut \& Mamajek 2013), although 
the hot spots associated with accretion will make the colors 
of PMS stars {\it bluer} than predicted by models. This being said, 
the characteristic timescale for the disruption of accretion disks in solar metallicity 
environments is on the order of 6 Myr (Haisch et al. 2001), and only a small fraction 
of any PMS stars in NGC 2401 would be expected to have accretion disks even 
with the young ages listed in Table 1. 

	If the models are assumed to reliably track PMS evolution 
then the comparison in the left hand panel of Figure 4 
might suggest that NGC 2401 contains a population of objects 
that have ages $< 10$ Myr. However, this is unlikely. First, if NGC 2401 contained 
a large population that is younger than 10 Myr then very massive hot MS stars might be 
expected in the cluster, and none are seen. Second, there should be a locus of PMS 
objects that belong to the older cluster population. If star formation occured in 
two discrete episodes then there should be two PMS sequences, and 
evidence for this is not seen in the CMDs.

	Binarity can have a profound influence on the location of objects on the CMD. 
A population of unresolved equal-mass binaries will fall 0.75 magnitudes above the 
single star sequence on CMDs. In fact, if the majority of 
cPMS stars are unresolved equal mass binaries then the agreement 
with the Padova isochrones is improved. This is demonstrated in the left hand 
panel of Figure 4, where the dashed red line shows the the 10 Myr model 
after it is shifted 0.75 magnitudes brighter. The shifted 10 Myr model now 
falls near the red envelope of the cPMS clump, while applying a similar shift to
the 32 Myr isochrone places it near the blue envelope of the cPMS clump. 
Still, this agreement requires NGC 2401 to have a very high frequency 
of unresolved equal mass binaries among objects with sub-solar masses, 
and this is contrary to what is seen in the field (e.g. Lada 2006). A modest 
population of wide binaries are found among the cPMS stars in these data (Section 4.2).

	There are uncertainties in the physics of PMS evolution, and these can lead 
to differences between evolutionary tracks produced with different codes. 
Evolutionary sequences from Siess, Dufour, \& Forestini (2000) are compared with 
the composite Rergion 1 -- 3 CMD in the left hand 
panel of Figure 5. The models were downloaded from 
the Siess web site \footnote[5]{http://www.astro.ulb.ac.be/~siess/WWWTools/Isochrones}, 
and use the transformation between effective temperature and magnitude
from Siess, Forestini, \& Dougados (1997). The downloaded isochrones are in the 
Cousins magnitude system, and were transformed into the SDSS photometric system 
using the Smith et al. (2002) relations. While there are differences between 
the Siess et al. (2000) and Bressan et al. (2012) isochrones, the former models still 
fall well blueward of the locus of cPMS stars in the NGC 2401 CMD.

	We close this section by examining the color distribution of the 
cPMS objects. The histogram distributions of the $g'-i'$ colors of objects with $i'$ 
between 21.0 and 22.0 are shown in the upper 
right hand panel of Figure 5. An excess number of red objects in Regions 1, 2, and 3 
is clearly seen. The difference between the color functions in 
the upper right hand panel of Figure 5 is shown in the 
lower right hand panel. The differential color distribution extends over 
1.5 magnitudes in $g'-i'$, with a prominent peak near $g'-i' \sim 3$. The colors of 
the 10 and 32 Myr Siess et al. (2000) sequences at $i'=21.5$ are also shown. 
The model colors overlap with the differenced color distribution, indicating that the 
models are consistent with at least some cPMS stars in NGC 2401. Still, 
the vast majority of these objects fall $\sim 0.2$ magnitudes in $g'-i'$ 
redward of the 10 Myr sequence and $\sim 0.5$ magnitudes redward of the 32 Myr models. 

\section{THE DISTRIBUTION OF CLUSTER STARS}

\subsection{The Projected Distribution of Resolved Objects}

	An image showing the projected distribution of stars with 
$i'$ between 16 and 17.5, which is a range that samples MS stars in NGC 2401, 
is shown in the left hand panel of Figure 6. The pixel intensities 
reflect number counts in $29 \times 29$ arcsec bins. This bin size was selected 
as a compromise between angular resolution and S/N ratio, and the pixel intensities 
were smoothed with a 1 pixel gaussian to further suppress noise. 
The radial distribution of pixel intensities is shown in Figure 
7, where foreground star contamination has been corrected statistically by 
subtracting number counts made in Regions 4 and 5. 
There is a pronounced central peak in the distribution of MS stars.

	An image showing the distribution of objects with $i'$ between 21 and 22 and 
$g'-i' > 2.0$, which is a range that samples cPMS stars in NGC 2401, is shown in the 
right hand panel of Figure 6. The radial distribution of pixel intensities is 
shown in Figure 7. The objects in this sample are clustered about the 
center of NGC 2401, suggesting that the majority are cluster members. 
It can also be seen from Figure 7 that the cPMS stars 
have a less peaky central distribution than the MS stars.

	A control sample was defined to examine the distribution of 
likely field stars in the same brightness range as the cPMS stars, and these 
are defined to have $i'$ between 21.0 and 22.0 and 
$g'-i' < 2.0$. The spatial distribution of these objects is not shown in Figure 6 
as there is little detail when displayed in image format. Indeed, the 
radial distribution of these objects -- shown in Figure 7 -- is indicative of a uniform 
distribution on the sky; the objects in the control sample thus do not follow the 
NGC2401-centered distributions of the MS and cPMS samples. This indicates that there is 
little or no contamination from cluster objects in this sample.

\subsection{The TPCF}

	The two-point correlation function (TPCF) probes the clustering properties 
of ensembles of objects. The TPCF is the distribution of separations for all possible 
pairings in a field, normalized to the separation function of a randomly distributed 
sample of objects having the same geometry as the science images -- geometric effects 
due to the shape and finite size of the detector are thus divided out. 
The TPCF multiplexes information in a manner that is systematically different from 
radial star counts, and so has the potential to reveal additional information 
about the distribution of objects in and around a cluster. A shortcoming 
is that positional information is not retained.

\subsubsection{Large Scale Clustering}

	The TPCFs of the MS and cPMS samples defined in Section 4.1 are 
shown in Figure 8. The y-axis shows the number of all 
possible pairings of objects in 15 arcsec separation bins, divided by 
the separation function of a randomly distributed sample of objects. This ratio was 
then normalized according to the number of all possible pairings to 
produce an amplitude spectrum that is independent of sample size. The TPCF of a uniformly 
distributed population will thus be flat with Power $= 1$. A 15 arcsec width for binning 
was selected as a compromise between angular resolution 
and the suppression of bin-to-bin noise. The random uncertainties at various 
separations can be estimated from the localised bin-to-bin jitter (Davidge 2012). 

	The TPCF of the MS sample shows a higher degree of clustering than that of the 
cPMS sample at separations $\leq 150$ arcsec, as expected given the greater 
central concentration of MS stars in Figures 6 and 7. The MS TPCF is not flat at 
large separations, hinting at organized structure among MS stars over much of the 
GMOS field. The slope of the MS TPCF changes near 250 arcsec, hinting at 
a change in the distribution of MS objects in the outer regions of the cluster. 
It thus appears that -- while cluster MS stars are centrally concentrated 
-- some cluster MS stars are present at distances of $250/2 = 125$ arcsec 
from the cluster center.

	The cPMS TPCF shows smaller amplitude variations than the MS TPCF. This is 
consistent with cPMS objects having a flatter radial distribution near the 
center of NGC 2401 than MS stars, as shown in Figures 6 and 7. In fact, the TPCF of cPMS 
objects is more-or-less flat at separations $< 120$ arcsec, suggesting a 
uniform distribution within $\sim 60$ arcsec of the cluster center. 
The degree of clustering amongst cPMS objects drops at separations $\geq 140$ 
arcsec, and the rate of decline is comparable to that in the MS TPCF. 
The cPMS TPCF flattens at separations $< 320$ arcsec, suggesting that field stars 
dominate the signal at these separations. 

	The difference between the cPMS and MS TPCFs is shown in 
the bottom panel of Figure 8. While there are significant 
departures from a horizontal line at separations $< 240$ arcsec, 
between 240 and 380 arcsec the difference between the TPCFs is a constant, suggesting 
that the clustering properties of the MS and cPMS samples at these separations are 
similar. This is consistent with the radial profiles of the cPMS and MS samples in Figure 
7, which have similar slopes at large radii. It should be recalled that the correlation 
function of the MS and cPMS samples are not flat in this separation  
interval, suggesting that the cluster extends beyond the flat region in the 
bottom panel of Figure 8.

	Non-cluster members -- which presumably are distributed uniformly 
across the GMOS science field -- will affect efforts to probe structure using the TPCF, 
especially in the diffusely populated outer regions of the cluster. The MS and 
cPMS samples have different fractional contamination from non-cluster objects, with 
the contamination among MS stars in Regions 2 and 3 amounting to $\sim 20\%$, while 
among cPMS stars field stars account for $\sim 27\%$ of sources. 
Davidge et al. (2013) examine the impact of field star contamination on the TPCF of 
Haffner 16 by adding stars with a uniform distribution to existing stellar catalogues. 
For the current study, a suite of TPCFs were generated for the MS sample after it had 
been supplemented with a uniform population of randomly positioned objects to simulate 
increasing the level of field star contamination to that in the cPMS sample. 
The MS TPCFs that contain a 7\% enhancement in field star numbers are practically 
indistinguishable from those generated using the real data, and so are not shown in 
Figure 8. These experiments suggest that while the fractional contamination from 
non-cluster sources in the MS and cPMS sample differs, this difference 
can not account for the distinct clustering properties 
of MS and cPMS stars across the GMOS field.

\subsubsection{Small Scale Clustering}

	The binary fraction is an important probe of cluster 
parameters and the dynamical state of a cluster (e.g. Bonatto et al. 2012). 
If the maximum separation between stars in binary systems is 0.04 pc 
(Larson 1995), then binaries in NGC 2401 will have separations $\leq 1.5$ arcsec. 
The TPCFs of objects in NGC 2401 with separation $< 4$ arcsec 
are shown in the Figure 8 insets.

	The MS and cPMS samples notionally contain cluster members 
within narrow brightness -- and hence mass -- ranges, and so 
physical binaries within these samples will have mass ratios near unity. 
In addition, given the mean masses of the cPMS and MS samples ($\sim 0.5$ vs.$\sim 
1.2$M$_{\odot}$ for cluster members) then the binary frequencies of these samples might 
be expected to differ by a factor of $\sim 2$ if in NGC 2401 (1) wide binaries faithfully 
track the binary frequency over all separations, and (2) the binary frequency 
is like that in the solar neighborhood (e.g. Figure 1 of Lada 2006). It can be seen 
from Figure 8 that the MS and cPMS TPCFs at small separations do differ. Taken at face 
value, the TPCFs in the insets suggest that the frequency of wide binaries 
among MS stars in NGC 2401 is $\sim 3\times$ higher than among PMS stars.

	Secular processes will produce differences in the 
binary frequencies of MS stars and PMS objects. Marks \& Kroupa (2012) 
investigate models in which clusters start with a common initial 
binary frequency. At a given separation lower mass PMS binaries 
are more susceptible to disruption than higher mass MS binaries, and so a lower 
binary fraction among PMS stars might be expected. Furthermore, if the mass spectrum 
within binary systems follows the IMF then the majority of all binary systems formed will 
have low mass companions. These low mass stars are ejected preferentially from binary 
systems, thereby forming a pool of companionless low mass objects (Marks \& Kroupa 2011).

	Physical binaries are almost certainly not the only cause 
of clustering signal in the TPCF at small separations. That a binary sequence is not 
apparent in the cluster CMD hints at a modest fraction of unresolved binaries, 
although contamination from non-cluster objects in the CMDs frustrates efforts to 
identify a binary sequence. The degree of clustering in the MS sample at small 
angular scales also has a broader distribution than expected for a maximum binary 
separation of 0.04 parsecs. While the orbits of some binary systems may have 
ellipticities that result in separations in excess of the 
nominal 0.04 pc value, such systems are likely to be rare as 
they will be the most susceptible to disruption by interactions with other 
cluster members. Contamination from foreground stars may contribute to the broadening 
of the peak in the MS TPCF at small separations, as many of the field stars 
are closer than NGC 2401. Given the relatively compact central core of 
bright MS stars then it is also likely that at least some of the 
clustering power may be due to chance superpositions in the two-dimensional projection of 
a three-dimensional structure. The peak in the MS power at small separations may 
also be a signature of sub-structuring among the MS stars in NGC 2401 
on $\sim 0.1$ parsec scales. 

\section{THE LUMINOSITY FUNCTION}

	Studies of the radial characteristics of stellar 
content provide insights into cluster evolution. 
The $i'$ LFs of sources in Regions 1, 2, and 3 are shown in 
Figure 9. The LFs in this figure were corrected statistically for non-cluster sources by 
subtracting the mean LF of objects in Regions 4 and 5. The LFs of Regions 2 and 3 
have been summed to boost the S/N ratio.

	The LF number counts were made with no color selection. While color selection 
has the potential to optimize the contrast with respect to background populations, the 
color properties of the target sample must be known in advance. Given that the 
dispersion in the properties of the cPMS stars in NGC 2401 are uncertain it 
was decided to adopt the widest possible color range for generating the LFs. Even with 
this conservative approach, the comparisons in Figure 9 suggest that 
objects with $i' > 21$ are present in statistically significant numbers in Regions 
1 -- 3 after correcting statistically for non-cluster sources. This is not unexpected 
given that the spatial distribution of objects in the cPMS sample (Section 4).

	The shape of the LF at the bright end changes with distance from the 
cluster center. The Region 1 LF is more-or-less flat at all but the faintest magnitudes, 
whereas there is a steady increase in number counts towards progressively 
fainter magnitudes at the bright end of the Region $2+3$ LF. Radial variations in 
stellar content are further examined in the bottom panel of Figure 9, where 
the difference between the Region 1 and the Region $2+3$ LFs -- after 
the former had been scaled to match the numbers of sources in 
the latter -- is shown. The difference in number counts is significant at the 
$2\sigma$ level near the bright end. However, the difference at the faint end 
is consistent with zero within the $1\sigma$ uncertainties. Thus, while cPMS 
and MS stars in NGC 2401 have different distributions on the sky, 
the shape of the LF at the faint end does not appear to change with radius.

	Davidge (2014) found that the $K$ LF of NGC 2401 
could be matched by models that assume a Chabrier (2001) MF. The Region 1 
and Region 2$+$3 $i'$ LFs are compared with models that adopt this MF in Figure 10. 
The models are constructed from the Z=0.020 Bressan et al. (2012) isochrones, and ages of 
20, 32, and 40 Myr are considered. The models have been scaled to match the observations 
in the normalization interval that is indicated. While the isochrones 
do not match the colors of the cPMS stars on the CMDs (Section 3), 
the model LFs do a reasonable job of matching the 
overall shape of the observed LFs at the faint end (see below). 

	The model LFs have low amplitude peaks at intermediate magnitudes that 
are due to the piling-up of PMS stars shortly before 
relaxing onto the MS. Cignoni et al. (2010) discuss the brightness of 
this bump as a potential age indicator for systems that are younger than 30 Myr. 
Unfortunately, the amplitudes of these bumps in the models is comparable to the 
$2\sigma$ dispersion in the NGC 2401 LFs, rendering 
as problematic efforts to measure the age of NGC 2401 from this feature. 
Still, these bumps might be useful for constraining the ages of young clusters 
that are at least $\sim 2 - 3 \times$ more massive than NGC 2401.

	The 20 Myr models are steeper than the Region 1 LF at 
magnitudes $i' > 19$, whereas the 32 and 40 Myr models more-or-less match 
the flat shape of the Region 1 LF when $i' < 20$. 
The faint end of the Region 1 LF is in better agreement with the 32 and 40 Myr models 
than the 20 Myr models. The reader is also cautioned that the statistical significance 
of the difference between the 32 and 40 Myr models and number counts at the faint end 
is not as high as the comparisons in Figure 10 might suggest. A shift of only 
0.05 magnitude in the adopted distance modulus to higher values would greatly 
reduce the disagreement with the number counts for $i' > 21$, while not affecting the 
agreement at the bright end.

	The Region $2+3$ LF is compared with the models 
in the middle panel of Figure 10. The agreement with all three models 
is poor for $i' < 18$, in the sense that the models overestimate the number of MS stars. 
As might be expected from the comparison in the lower panel of 
Figure 9, the difference between the models and the Region $2+3$ LF at the faint 
end is similar to that found for the Region 1 LF. 

	The sum of the LFs in all three regions is compared with the models in 
the bottom panel of Figure 10. The agreement with the models at 
the bright end is slightly better than for the Region 1 LF alone. 
Still, the 20 Myr model predicts number counts at $i' = 18.5$ that differ 
from the observations at almost the $3\sigma$ level. 
The number counts with $i' < 20$ are better matched by 
the 32 and 40 Myr models than by the 20 Myr model.

	Recognizing that even modest uncertainties in the 
distance modulus will affect the ability of models to 
match the faint end of the LF, then the comparisons in the bottom panel of Figure 10 
are consistent with a Chabrier (2001) MF matching the LF of NGC 2401 over a 
magnitude range that samples both MS and PMS stars. To the extent that the Chabrier MF 
is universal -- and so provides a benchmark for gauging mass loss from a cluster -- 
then even though the distributions of bright and faint objects suggest that there is mass 
segregation in NGC 2401 (Section 4), the cluster has apparently not yet lost significant 
numbers of objects with $i' < 21 - 21.5$, which corresponds 
to masses $\sim 0.5$M$_{\odot}$. 

\section{SUMMARY \& DISCUSSION}

	Deep $g'$ and $i'$ images recorded with GMOS-S have been used to investigate the 
photometric properties of stars in the open cluster NGC 2401. 
NGC 2401 is at a distance of 5 kpc in the direction of the Galactic anticenter, 
placing it in the Perseus spiral arm well outside the Solar circle. To the extent that 
environment plays a role in defining the timescale of cluster destruction (e.g. de Grijs 
et al. 2013), then clusters at large Galactocentric radii like NGC 2401 may be 
useful laboratories for examining sub-structures imprinted early in cluster evolution, 
that may be erased early-on in higher density parts of the Galaxy.

	The CMD extends to substantially fainter magnitudes than previous studies, 
sampling much of the MS as well as cPMS objects with masses as low as 
$\sim 0.5$M$_{\odot}$, which is the approximate faint limit of these data. 
The objects that are identified as cPMS objects are too red to be 
MS stars in NGC 2401, and have a spatial distribution 
that is centered on the cluster. The cPMS stars also occur in numbers that are 
roughly consistent with those expected for PMS objects in young clusters (Section 5). 

	MS stars and cPMS objects in NGC 2401 
have different spatial distributions near the cluster center, with the cPMS stars 
having a less peaked radial distribution than the more massive MS stars. 
Evidence to support this comes from the projected distribution of objects 
on the sky (Section 4.1), the TPCF (Section 4.2), and the cluster LF (Section 5). This 
difference in distributions was not detected by Davidge (2014) because of the small 
GeMS field size.

	The LF of NGC 2401 can be fit by a model that assumes a Chabrier MF, as might 
be expected if the universality of the IMF holds over a wide range of cluster masses 
(e.g. Hennebelle 2012). If the IMF is universal then the agreement with 
the Chabrier MF suggests that NGC 2401 has not yet lost large numbers of stars with 
masses $\geq 0.5$M$_{\odot}$. The MF constructed by Baume et al. (2006), which is 
shown in their Figure 11 and assumes that all stars are on the MS, 
breaks downward when $\leq 1$ M$_{\odot}$. If NGC 2401 has an age near 
25 Gyr then the Bressan et al. (2012) models predict that the 
MSCO occurs at ZAMS masses $\sim 1.2$M$_{\odot}$. 
This raises the possibility that the break in the Baume et al. (2006) MF is 
due to the MSCO, rather than a change in the character of the NGC 2401 MF. 

	A lower limit to the stellar mass of NGC 2401 can be estimated from the 
LF constructed in Section 5. The application of a mass-luminosity 
relation from the Padova models to the LF in the bottom panel 
of Figure 10 suggests that $\sim 600$ M$_{\odot}$ is sampled with these 
data. Given that the faint limit of the LF corresponds to objects with 
masses $\sim 0.5$ M$_{\odot}$ then, depending on the shape of the MF at the lower mass 
end, the total stellar mass is likely much larger than this. NGC 2401 appears to be 
sufficiently massive that there is a reasonable chance that at least one star more 
massive than 10M$_{\odot}$ formed. The presence (or absence) of such massive stars 
can affect the stellar content and overall evolution of low mass clusters 
(Pelupessy \& Portegies Zwart 2012). As voracious consumers of star-forming material, 
the most massive stars may form early-on (Kirk et al. 2014), thereby increasing 
the likelihood that they will affect the gas content of a star-forming region 
before large numbers of lower mass stars have formed. 

	The differences in the projected distributions of 
MS and cPMS stars indicate that there is mass segregation in NGC 2401. 
It is unlikely that this segregation is the result of two-body relaxation, as 
the Region 1 LF does not show evidence of a deficiency in cPMS objects, while 
the entire cluster LF can be modelled with a solar 
neighborhood-like MF. Another possibility is that the 
differences in the distributions of MS and cPMS objects might be the consequence of 
large-scale mass loss early in the life of a cluster. The response of a cluster to the 
sudden loss of material depends on its dynamical state at the time of mass ejection and 
the presence of dense sub-structures where the star-forming efficiency has been high 
(e.g. Smith et al. 2011). If NGC 2401 was in a partially virialized 
expanding state at the time that significant mass loss occured 
(presumably within the first few Myr years of its life 
when SNe and/or stellar winds would be most active) then lower mass members 
-- which would have higher mean velocities than the more massive MS stars -- would 
move to larger radii in response to the change in the depth of the gravitational 
potential. However, at large radii the distributions of MS and cPMS stars in 
NGC 2401 are very similar (e.g. Figure 7).

	The relative distributions of MS and cPMS stars might be the result 
of primordial sub-structuring that has not yet been erased. Sub-structuring is 
seen in clusters that are as old as 100 Myr (e.g. Sanchez \& Alfaro 2009). 
Indeed, Santos-Silva \& Gregorio-Hetem (2012) find that a 
large fraction of the nearby young clusters in their sample 
are hierarchical systems with multiple peaks in their light profiles. 

	The young star complex in the Orion Nebula may provide clues to the 
relative distributions of MS and cPMS objects in NGC 2401. Hillenbrand \& Hartmann 
(1998) investigated the spatial distribution of objects throughout the 
Orion complex, and found evidence of mass segregation. They argue that the young age 
of this region makes secular processes an unlikely cause of mass segregation, 
and suggest that the observed stellar distribution could result 
if higher mass stars tend to form in the densest regions of 
a protocluster, as this is where there is the highest probabilty of massive 
star formation due to the ready access to material that can be accreted
(e.g. Bonnell et al. 1997). Still, such primordial sub-structuring is likely not a 
panacea for explaining all of the structural properties of NGC 2401. While 
simulations of very low mass systems with long relaxation times produce a 
central concentration of massive stars, in more massive 
systems with shorter dynamical time scales the final structure 
may reflect a mix of primordial and secular processes (e.g. Kirk et al. 2014). 

	Future work on NGC 2401 might involve obtaining deeper photometry 
that will plumb lower mass cluster members. If the MF follows that in the 
solar neighborhood then very low mass PMS stars should form a conspicuous sequence 
on CMDs - the failure to detect such a sequence would be one indicator that the 
cluster has lost these objects. While the spatial distribution 
of cPMS objects with masses $\geq 0.5$M$_{\odot}$ shows no signs of an 
obvious tidal tail (Figure 6), objects with lower masses may be traced over 
even larger areas than explored here, and may reveal signs of dispersal from the 
cluster. Spectra of faint objects in and around NGC 2401 would also be useful to better 
distinguish between cluster and non-cluster members. Such observations would allow the 
distribution of PMS stars belonging to NGC 2401 to be assessed free 
of contamination from non-cluster objects. 

\acknowledgements{It is a pleasure to thank Doug Johnstone and Helen Kirk for interesting 
and productive discussions regarding the distribution of stars and protostars in young 
star-forming regions. Thanks are also extended to the anonymous referee for 
comments that improved the presentation of the results. This research has made use 
of the WEBDA database, operated at the Department of Theoretical Physics and 
Astrophysics of the Masaryk University.}

\appendix

\section{COMPARING $g'-i'$ and $V-I$ COLORS}

	The use of the Smith et al. (2002) transformation 
equations means that only those points in the Baume et al. (2006) 
dataset that have photometry in B, V, R, and I are included in 
Figure 3. A sample that extends to fainter magnitudes 
can be obtained if objects in the Baume et al. 
(2006) study that have only V and I photometry 
are considered. Here, the Baume et al. (2006) $V-I$ colors 
are compared with the $g'-i'$ colors obtained with GMOS. A matching radius of 0.7 
arcsec (i.e. 5 GMOS pixels) was adopted when pairing stars in the two datasets.

	The difference between the two colors is shown in the right hand panel of 
Figure A1. The left hand panel of Figure A1 shows the GMOS $(i', g'-i')$ CMD of 
objects that could be matched with Baume et al. (2006) observations. 
There is a clear absence of stars with $i' > 21.5$ and $g'-i' > 3$.
It can be seen in the right hand panel that both the mean color 
difference and the dispersion about the mean changes with 
$i'$. Applying an iterative $2.5\sigma$ clipping routine to suppress outliers, 
the mean color difference of objects with $i'$ between = 17.5 and 18.5 
is $-0.13$ magnitude with a standard devation of 
$\pm 0.05$ magnitudes. Between $i' = 20$ and 21 the mean color difference is 
0.11 magnitudes, with a standard deviation of $\pm 0.11$ magnitudes. A trend of larger 
differences between $g'-i'$ and $V-I$ towards redder colors is seen among 
Galactic disk stars (e.g. Table 3 of Jordi et al. 2006). The tendency for the color 
difference to increase towards fainter magnitudes in Figure A1 is thus to be 
expected given that red stars constitute a larger fraction of objects at faint 
magnitudes in NGC 2401 than at bright magnitudes.

\clearpage

\begin{table*}
\begin{center}
\begin{tabular}{llll}
\tableline\tableline
Age & Distance & E(B--V) & Reference \\
(Myr) & (kpc) & & \\
\tableline
1000 & $3.5 \pm 0.4$ & 0.0 & Sujatha et al. (2004) \\
$25 \pm 5$ & $6.3 \pm 0.5$ & $0.36 \pm 0.01$ & Baume et al. (2006) \\
100 & 5.6 & 0.24 $^1$ & Hasegawa et al. (2008) \\
$25 \pm 5$ & $5.5 \pm 0.1$ & $0.15 \pm 0.02$ & Davidge (2014) \\
\tableline
\end{tabular}
\caption{Cluster Parameters}
$^1$ Computed from E(V--I)$= 0.3$ using the reddening curve in 
Table 6 of Schlegel et al. (1998).
\end{center}
\end{table*}

\clearpage

\begin{figure}
\figurenum{1}
\epsscale{1.00}
\plotone{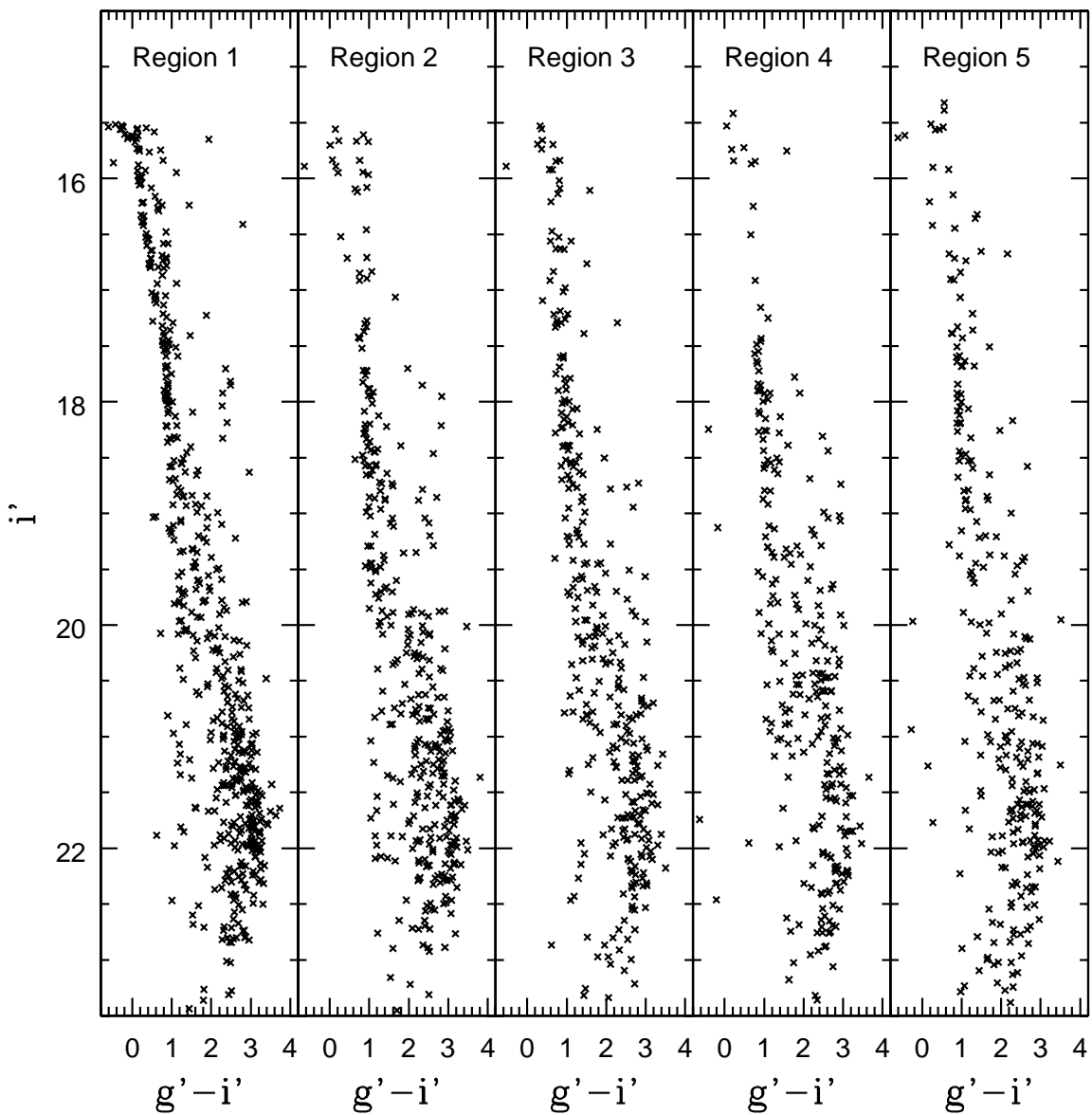}
\caption{Color-magnitude diagrams. The CMDs sample equal area regions 
that are centered on the main concentration of bright stars in NGC 2401. 
The photometry of objects with $i' < 17.5$ is from the short exposure 
images, and the bright limit is defined by detector saturation. 
Number counts suggest that Regions 1 -- 3 contain objects that belong to 
NGC 2401, whereas Regions 4 and 5 contain few cluster sources.}
\end{figure}

\clearpage

\begin{figure}
\figurenum{2}
\epsscale{1.00}
\plotone{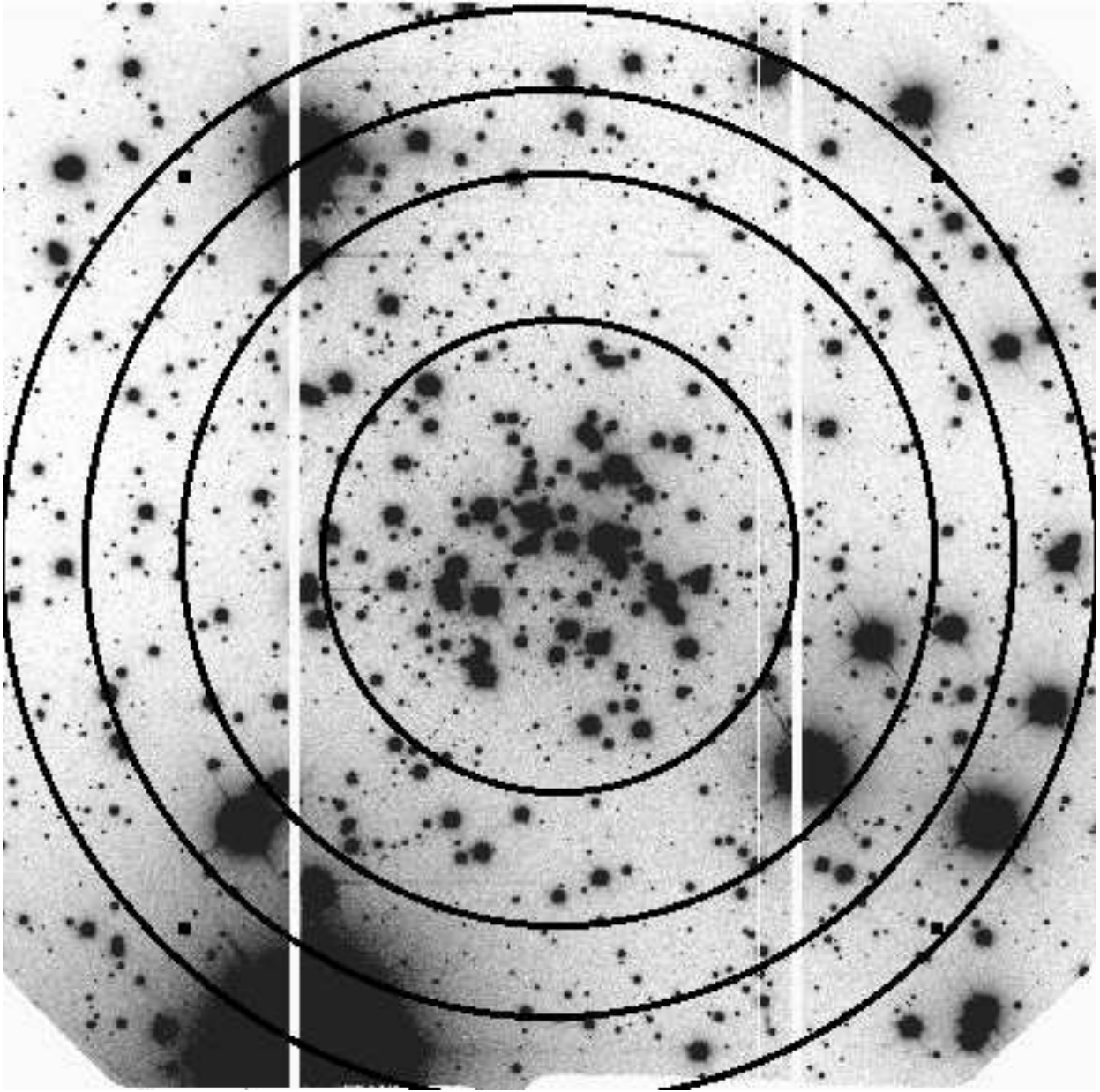}
\caption{Region boundaries. Region 5 encompasses the area that 
is external to the largest annulus. The backdrop is the long exposure 
$g'$ GMOS image, which covers $5.5 \times 5.5$ arcmin$^2$. 
North is at the top, and east is to the right. 
The two prominent vertical stripes are gaps between CCDs.}
\end{figure}

\clearpage

\begin{figure}
\figurenum{3}
\epsscale{1.00}
\plotone{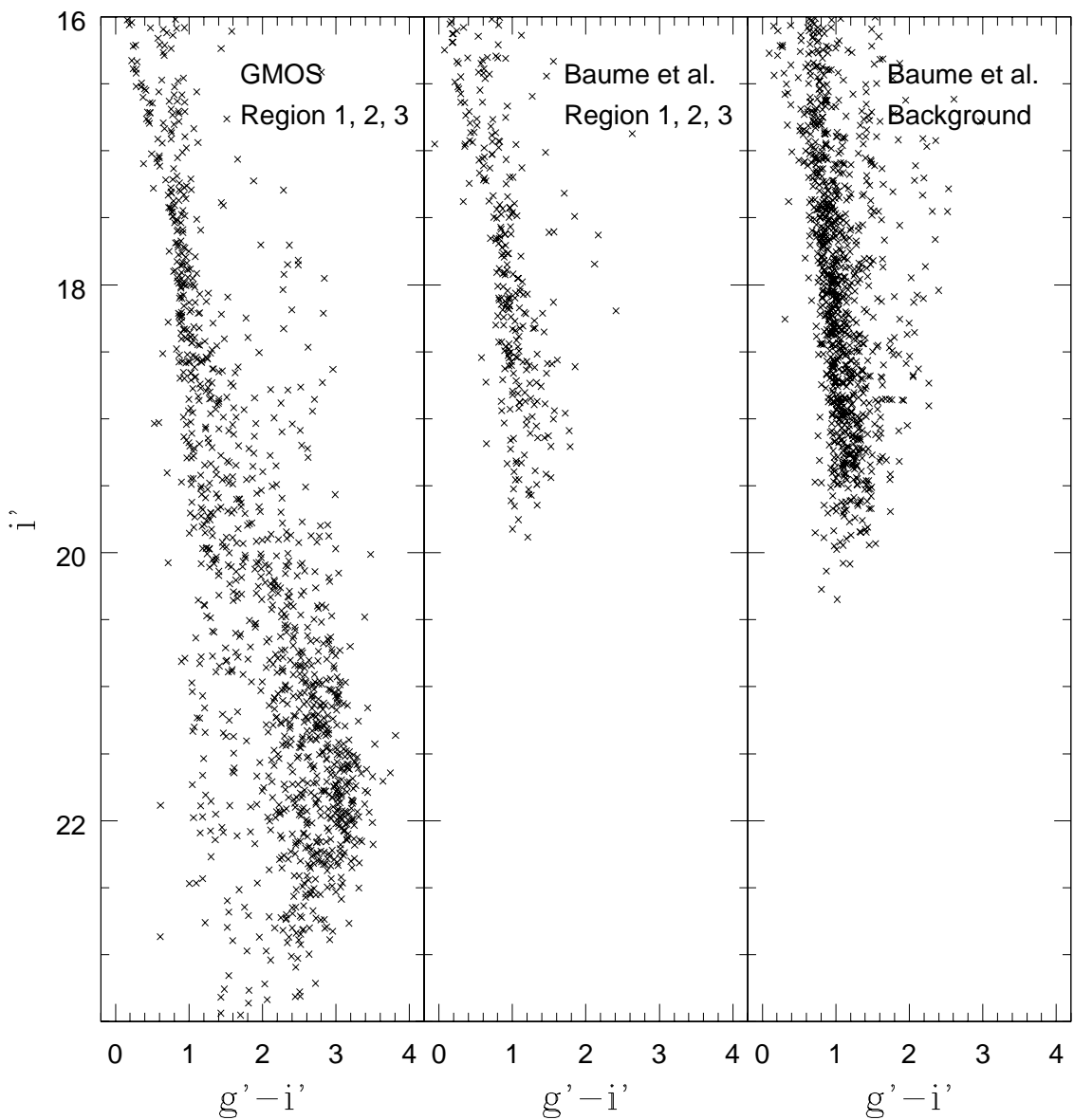}
\caption{Comparison with Baume et al. (2006) photometry. The composite CMD of Regions 1, 
2, and 3 obtained from the GMOS measurements is shown in the left hand panel. The CMDs 
of objects from Baume et al. (2006) that are in Regions 1, 2, and 3 (middle panel) and 
at larger radii (right hand panel) are also shown. The Baume et al. (2006) measurements 
were transformed into the SDSS photometric system using the 
relations in Table 7 of Smith et al. (2002).}
\end{figure}

\clearpage

\begin{figure}
\figurenum{4}
\epsscale{1.00}
\plotone{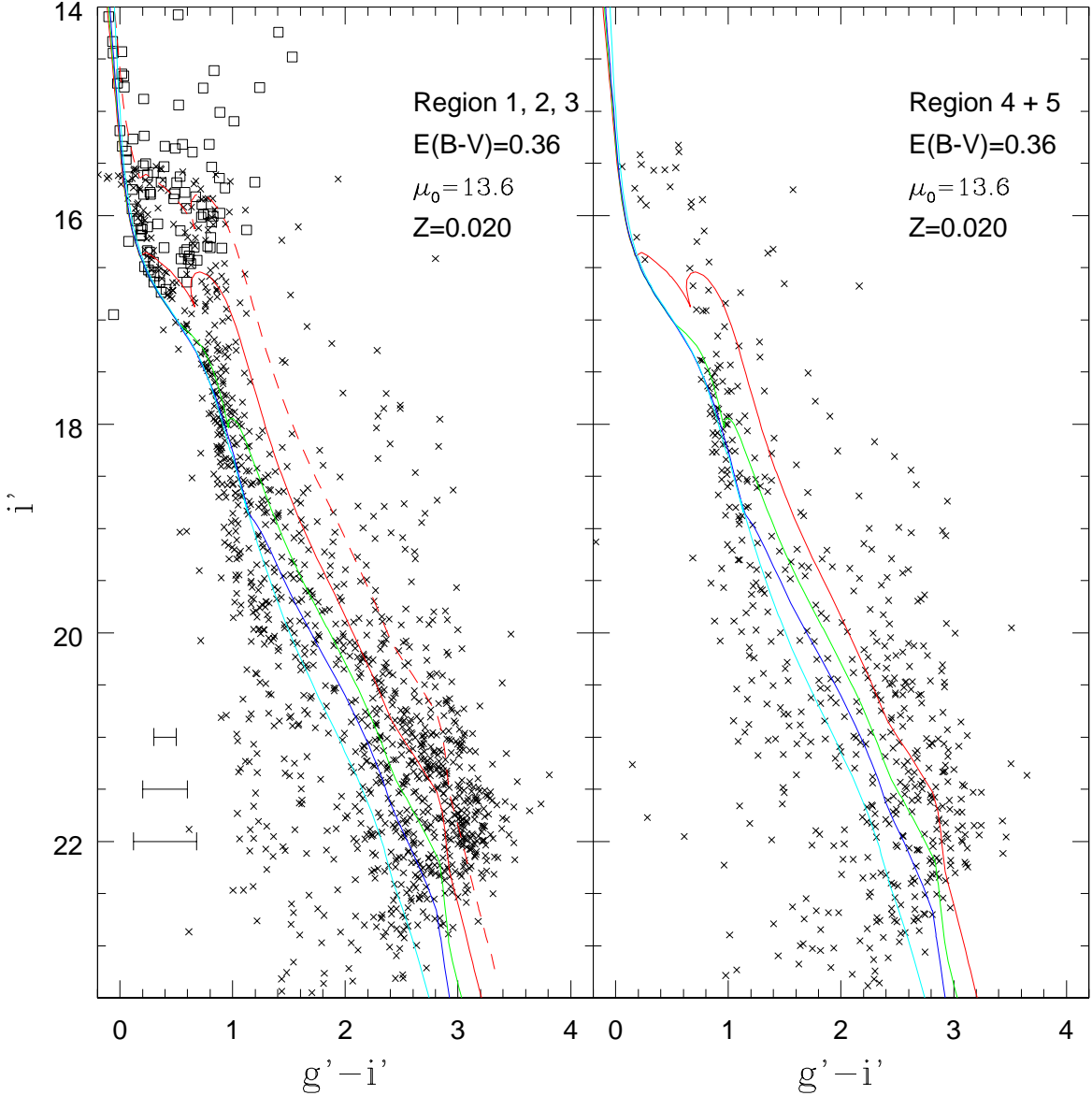}
\caption{Comparisons with isochrones. The $(i', g'-i')$ CMDs of Regions 1, 2, and 3 
(left hand panel) and Regions 4 and 5 (right hand panel) are shown. Regions 1, 2, and 3 
together sample an area that is 50\% larger than that covered by Regions 4 and 5. The 
error bars in the left hand panel show the $\pm 2\sigma$ dispersion in $g'-i'$ 
of sources with $g'-i' \sim 3$ obtained from artificial star experiments. The 
open squares are photometric measurements from Baume et al. (2006) that have been 
tranformed into the SDSS system. Isochrones from Bressan et al. (2012) that have 
Z=0.020 and ages of 10 (red), 20 (green), 32 (blue), and 71 (cyan) Myr are shown. The 
adopted distance modulus is 13.6, and the color excess is that measured by Baume et al. 
(2006). The dashed red line is the 10 Myr sequence shifted to brighter magnitudes by 
0.75 magnitudes to simulate a sequence of unresolved equal-mass binaries.}
\end{figure}
 
\clearpage

\begin{figure}
\figurenum{5}
\epsscale{1.00}
\plotone{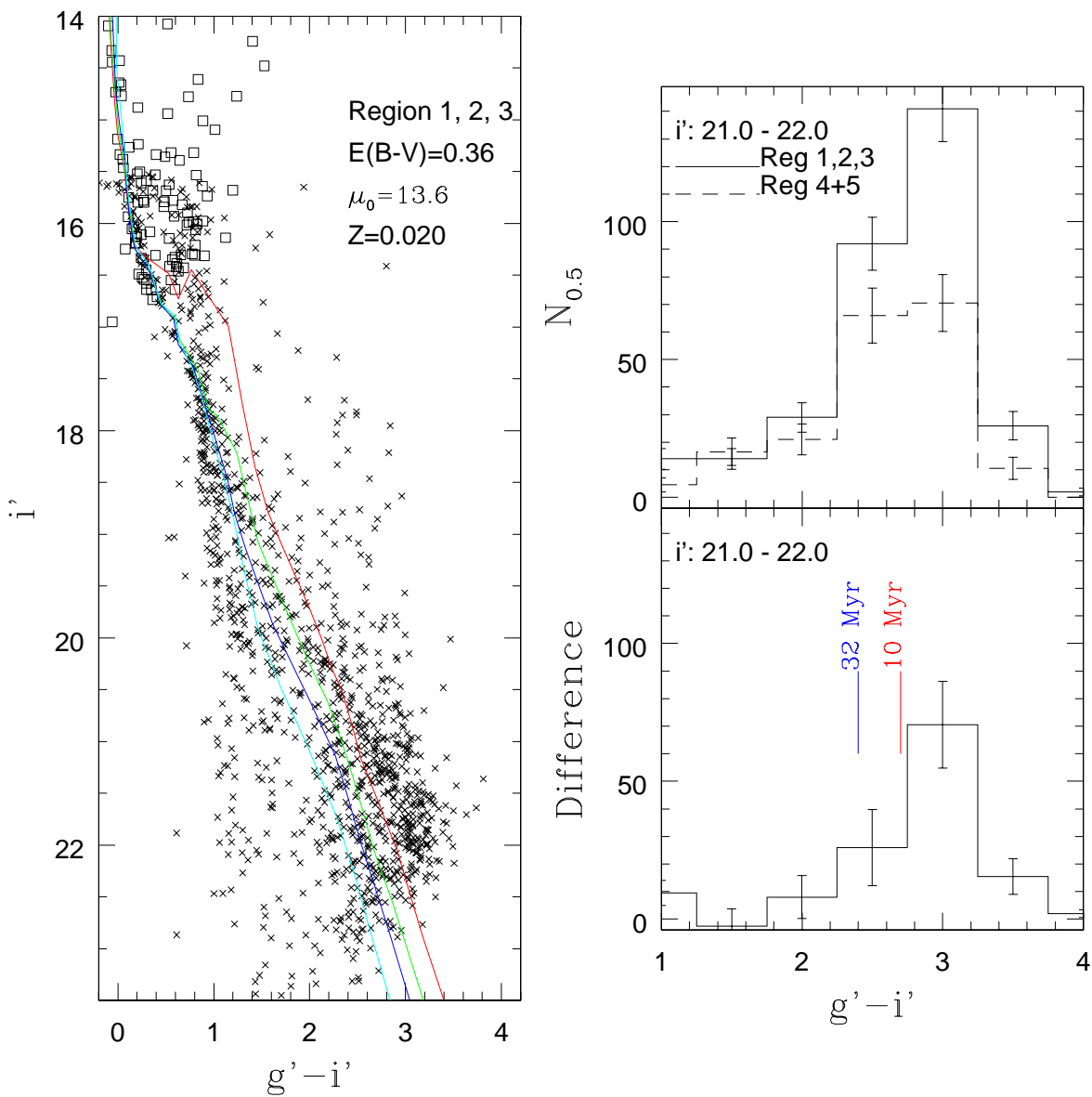}
\caption{Comparisons with isochrones from Siess et al. (2000). The CMD in the left hand 
panel is the same as that in the left hand panel of Figure 4, but with isochrones from 
Siess et al. (2000). The isochrones have the same ages and color coding as 
in Figure 4. The upper right hand panel shows the $g'-i'$ 
distribution of sources with $i'$ between 21.0 and 22.0 in Regions 1, 2, and 3 
(solid line) and Regions 4 and 5 (dashed line); the latter has been scaled 
to match the areal coverage of Regions 1, 2, and 3. The error bars show $\pm 1\sigma$ 
uncertainties calculated from counting statistics. The difference between these 
distributions is shown in the lower right hand panel. The $g'-i'$ colors of the 
10 and 32 Gyr Siess et al. (2000) isochrones at $i'=21.5$ are indicated.}
\end{figure}
 
\clearpage

\begin{figure}
\figurenum{6}
\epsscale{1.00}
\plotone{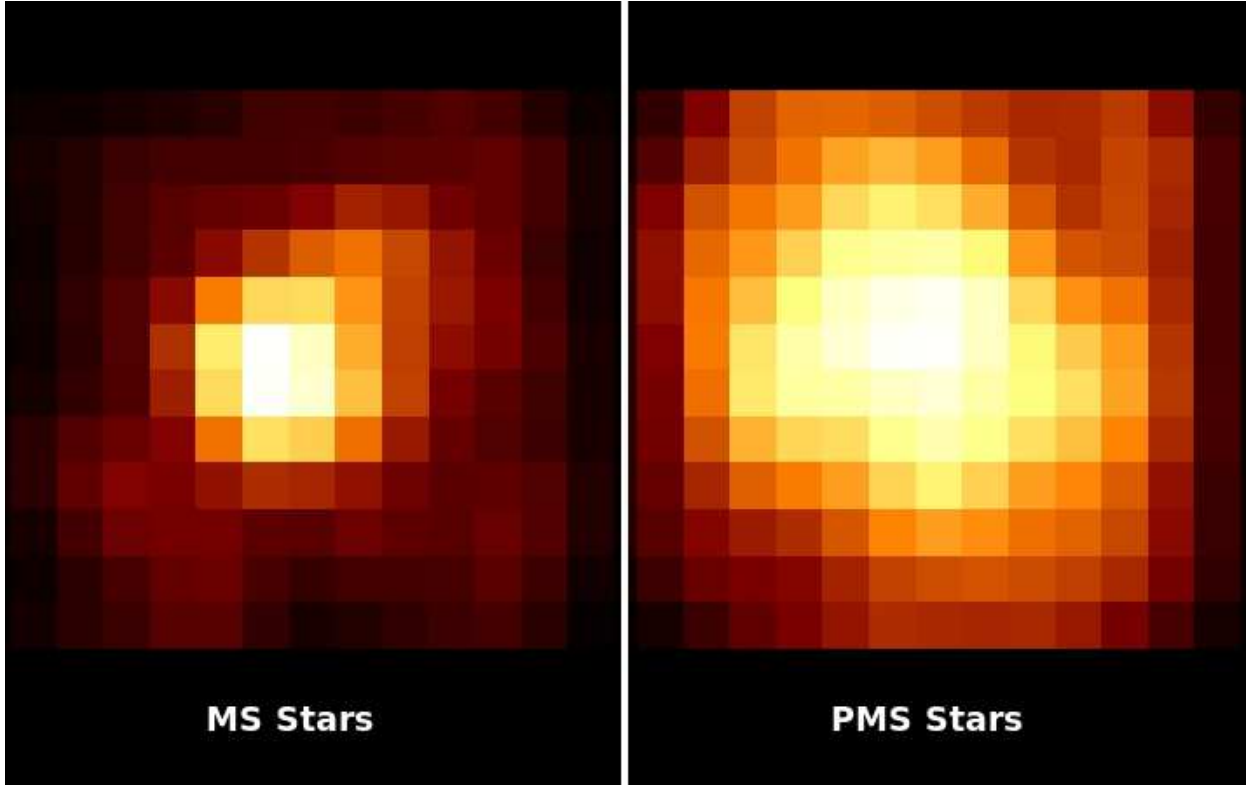}
\caption{Projected distributions of photometrically-selected 
samples. The panels show the locations of objects with $i'$ 
between 16 and 17.5 (MS Stars; left hand panel) and $i'$ between 
21 and 22 and $g'-i' > 2$ (cPMS Stars; right hand panel). North is at the top, and East 
is to the right. The pixel intensities show the number of objects in $29 \times 29$ 
arcsec bins after normalization to the total number of objects in each sample and 
convolution with a $\sigma = 1$ pixel Gaussian. The images are displayed with the 
same stretch.}
\end{figure}

\clearpage

\begin{figure}
\figurenum{7}
\epsscale{1.00}
\plotone{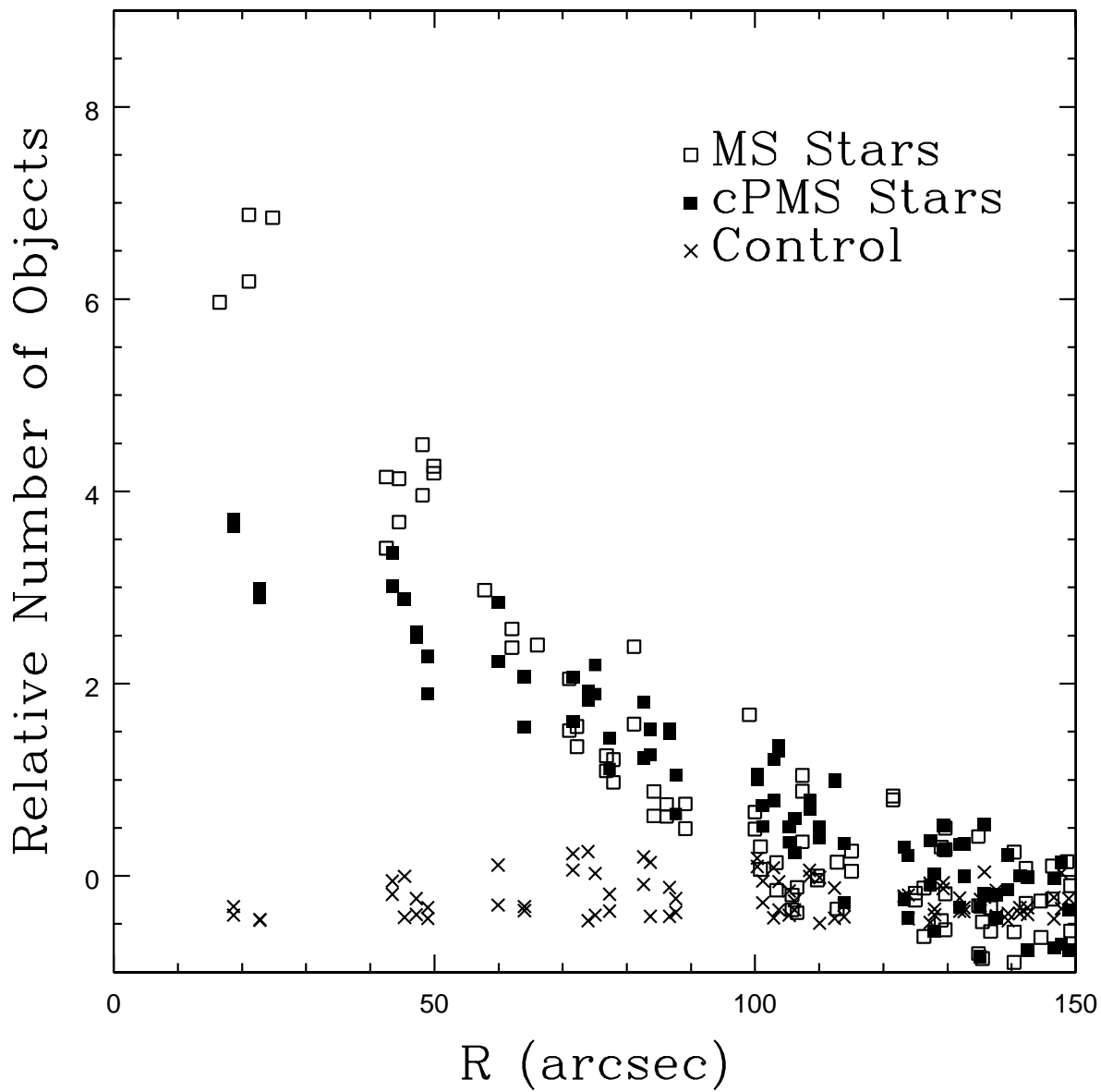}
\caption{Radial profiles of the images in Figure 6. The solid squares 
show number counts for the cPMS sample, while the open 
squares show number counts for the MS stars. The profiles 
have been corrected for field star contamination by subtracting number counts 
made in Regions 4 and 5, and the results have been scaled to balance the total number 
of sources in each sample. The distribution of MS stars has a prominent 
central cusp, whereas the cPMS stars are not as centrally concentrated. The 
crosses show the radial distribution of objects that have the 
same brightness range as cPMS objects, but bluer colors. These sources have 
a more-or-less uniform on-sky distribution, indicating little or no contamination 
from cluster stars in this brightness and color range.}
\end{figure}

\clearpage

\begin{figure}
\figurenum{8}
\epsscale{1.00}
\plotone{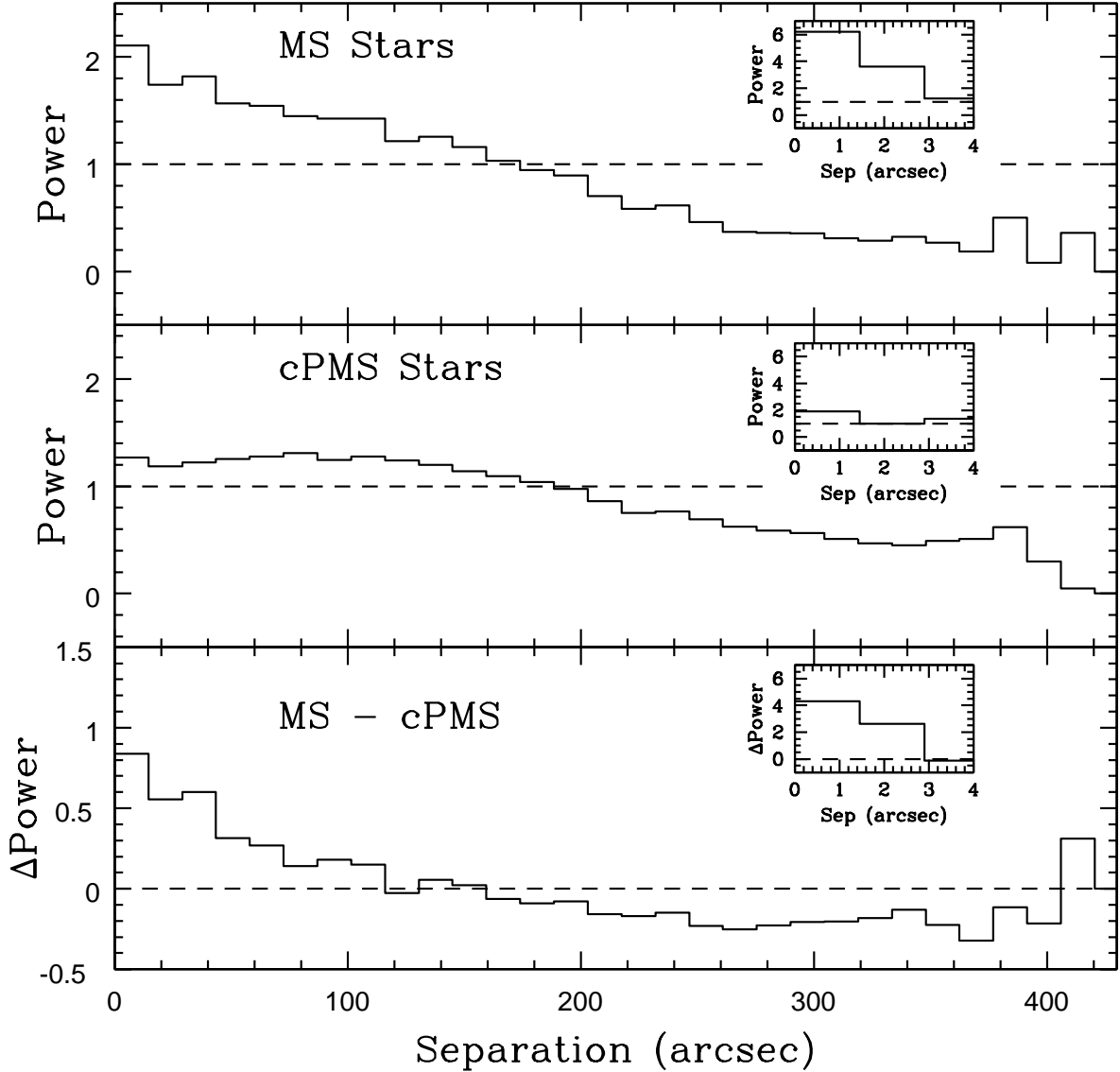}
\caption{Correlation functions. The y-axis shows the number of all 
possible pairings of objects per 15 arcsec separation interval, divided by 
the separation function of a randomly distributed ensemble of stars and scaled to 
the total number of pairings. The TPCF of a uniformly 
distributed population then has a uniform power $= 1$. Correlation 
functions with finer angular binning and restricted to separations $< 4$ arcsec, 
which is the angular range where binaries and compact sub-structures will be resolved 
in NGC 2401, are shown in the insets. The difference between the TPCFs of the MS and cPMS 
samples is shown in the bottom panel.}
\end{figure}

\clearpage

\begin{figure}
\figurenum{9}
\epsscale{0.80}
\plotone{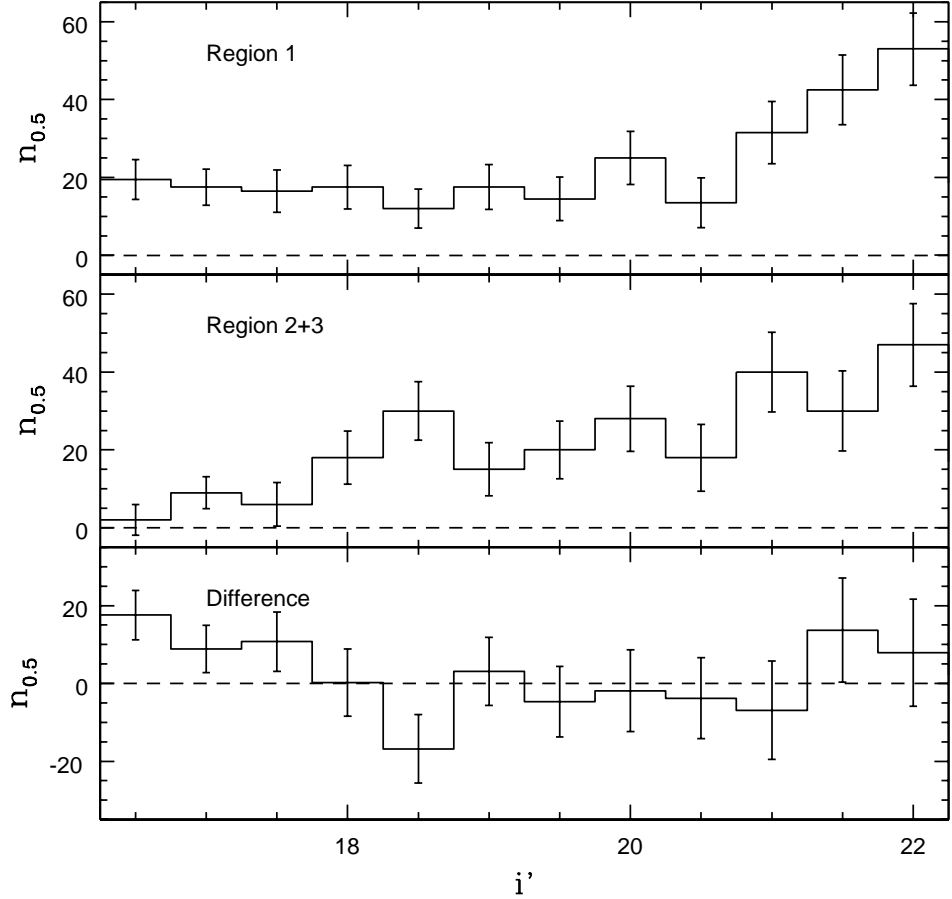}
\caption{Luminosity functions. n$_{0.5}$ is the number of objects per 0.5 
magnitude interval in $i'$ corrected statistically for non-cluster contamination. 
The Region 2 and 3 LFs are summed in the middle panel to 
boost the S/N ratio. The dashed line indicates zero counts. 
The error bars show $1\sigma$ uncertainties computed from Poisson statistics. The 
difference between the Region 1 and Region $2+3$ LFs -- with the latter scaled to match 
the number of objects in the former  -- is shown in the bottom panel. Statistically 
significant departures from zero in the lower panel occur at the bright end 
in the magnitude interval that contains MS stars $(i' < 17.5)$, 
further confirming the centrally-peaked distribution of bright MS stars.}
\end{figure}

\clearpage

\begin{figure}
\figurenum{10}
\epsscale{0.80}
\plotone{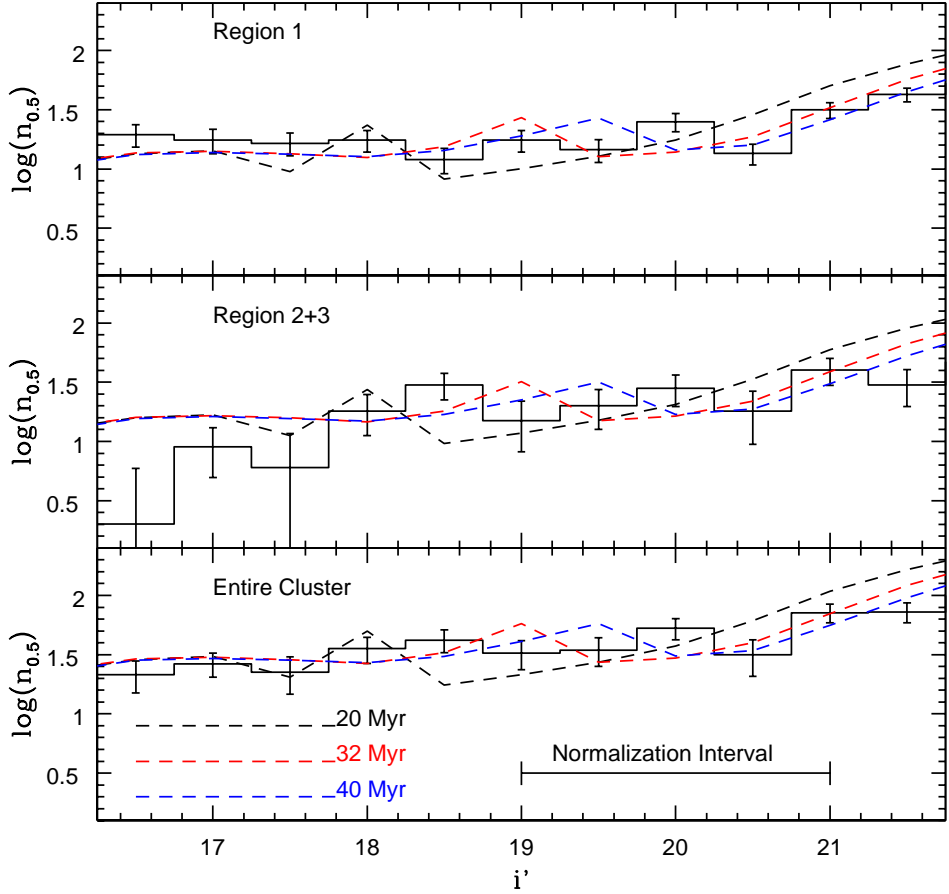}
\caption{Comparisons with model LFs. The LFs in the top and middle panels are 
from Figure 9, and the error bars show $\pm 1\sigma$ uncertainties. 
Artificial star experiments suggest that the data are complete in the magnitude 
interval shown. The dashed lines are Z=0.020 LFs constructed from the Bressan et 
al. (2012) models for a population with a Chabrier (2001) MF. The models have been 
scaled to match the observations in the marked magnitude interval, and 
have ages of 20 Myr (black), 32 Myr (red), and 40 Myr (blue). Peaks that reflect the 
build up of PMS stars on the CMD prior to relaxing onto the MS are seen in 
all three models.}
\end{figure}

\clearpage

\begin{figure}
\figurenum{A1}
\epsscale{0.80}
\plotone{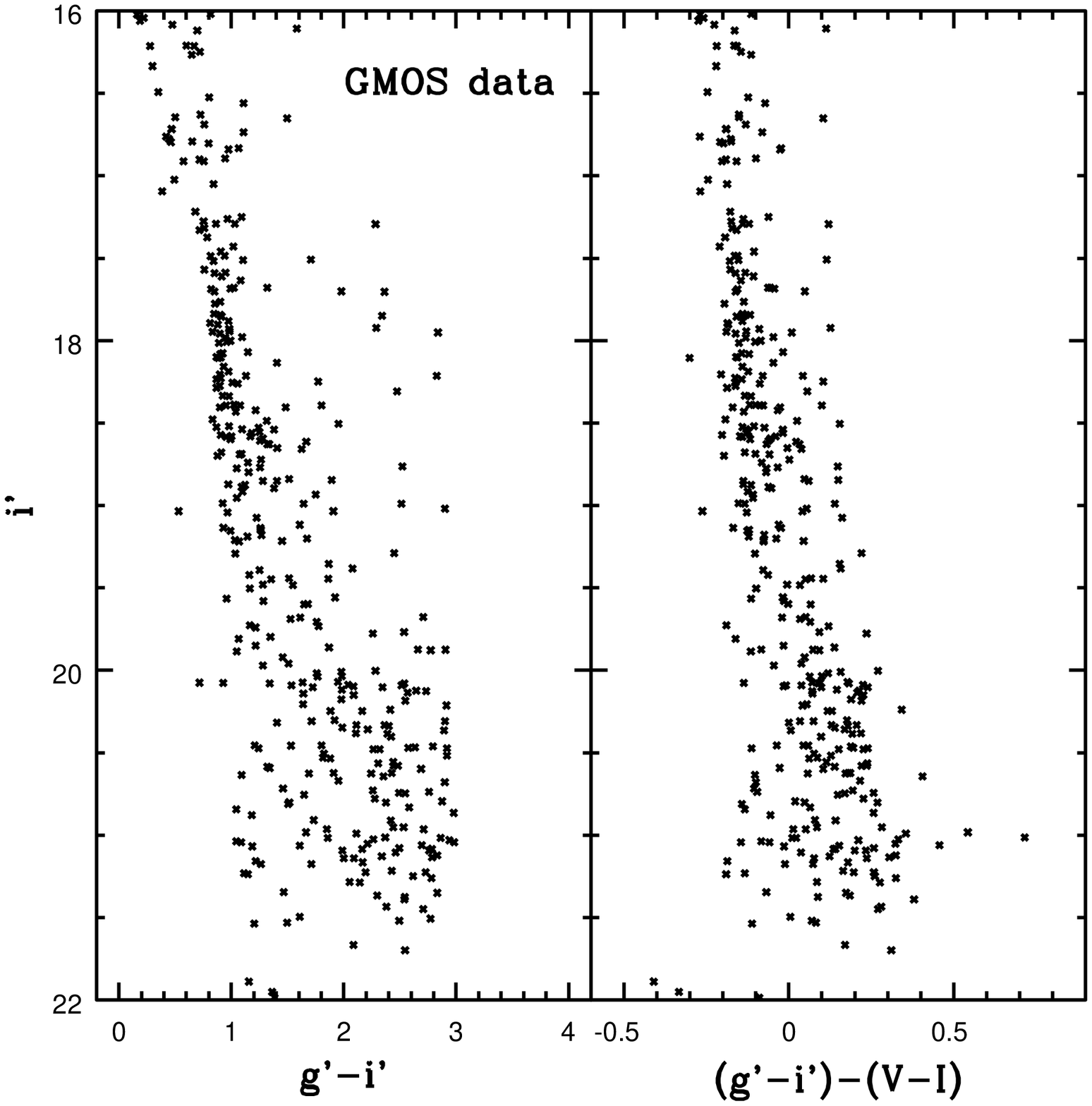}
\caption{Comparing $g'-i'$ and $V-I$ colors. The left hand panel shows the GMOS $(i', 
g'-i')$ CMD of objects that were matched with sources in the 
Baume et al. (2006) dataset using a 0.7 arcsec matching radius. The right hand panel 
shows the difference between the $g'-i'$ and $V-I$ colors. The $1\sigma$ dispersion 
in $(g'-i')-(V-I)$ is $\pm 0.05$ magnitudes near $i = 18$, and $\pm 0.11$ 
magnitudes near $i' = 20.5$. The tendency for larger $(g'-i')-(V-I)$ values 
at magnitudes $i' > 20$, where red stars contribute more to the color 
distribution than at brighter magnitudes, is consistent with the measured $g'-i'$ 
and $V-I$ colors of stars in the solar neighborhood (e.g. Jordi et al. 2006).}
\end{figure}

\end{document}